\documentclass[useAMS,usenatbib]{mn2e}

\usepackage{graphicx}

\title[GR\,Mus]{Kinematical Studies of the Low Mass X-Ray Binary 
GR\,Mus (XB\,1254--690)}
\author[A. D. Barnes et al.]{A. D. Barnes$^{1}$\thanks{E-mail:
adb@astro.soton.ac.uk (ADB)}, J. Casares$^{2}$, R. Cornelisse$^{1}$, P. A. Charles$^{1,3}$, 
D. Steeghs$^{4,5}$,  
\newauthor R.I. Hynes$^{6}$ and K. O'Brien$^{7}$\\
$^{1}$School of Physics and Astronomy, University of Southampton, SO17 1BJ, 
UK\\
$^{2}$Instituto de Astrofisica de Canarias, 38200 La Laguna, Tenerife, Spain\\
$^{3}$South African Astronomical Observatory, PO Box 9, Observatory 7935, 
Cape, South Africa\\
$^{4}$Harvard-Smithsonian Center for Astrophysics, MS-67, 60 Garden Street, 
Cambridge, MA 02138, USA\\
$^{5}$Department of Physics, University of Warwick, Coventry, CV4 7AL, UK\\
$^{6}$Department of Physics and Astronomy, 202 Nicholson Hall, Louisiana State University, Baton Rouge, LA 70803, USA\\
$^{7}$European Southern Observatory, Casilla 19001, Santiago 19, Chile\\
}
\begin{document}

\date{}

\pagerange{\pageref{firstpage}--\pageref{lastpage}} \pubyear{2002}

\maketitle

\label{firstpage}

\begin{abstract}
We present simultaneous high-resolution optical spectroscopy and X-ray data of 
the X-ray binary system GR\,Mus (XB\,1254--690), obtained over a full range of 
orbital 
phases.  The X-ray observations are used to re-establish the orbital ephemeris 
for this source.  The optical data includes the first spectroscopic detection 
of the donor star in this system, through the use of the Doppler Tomography 
technique on the Bowen fluorescence blend ($\sim$4630-4650 \AA).  In 
combination with an estimate 
for the orbital parameters of the compact object using the wings of the 
He\,{\sc ii}~$\lambda$4686 emission line, dynamical mass constraints of 
1.20 $\leq M_X/M_{\sun} \leq$ 2.64 for the neutron star and 0.45 $\leq M_2/M_{\sun} \leq$ 0.85 for the companion are derived.

\end{abstract}


\section{Introduction}

XB\,1254--690 is a persistently bright, low-mass X-ray binary (LMXB).  
It was identified with a faint blue star (GR\,Mus, {\it V} = 19.1), which 
exhibited the Bowen blend of N\,III and C\,III in emission \citep{grif}.  The 
object produced Type 1 X-ray bursts \citep{mason,courv}, indicating 
the presence of a neutron star as the compact object.  
Dips of up to 95\% of the 1--10 keV flux, with a recurrence period of 
{\it $P_X$} = 3.88 $\pm$ 0.15 hr and lasting $\sim$ 0.8 hr per cycle, 
were discovered in {\it EXOSAT} data by \citet{courv}.  
The dips are caused by obscuration of the central source by a bulge on the 
outer edge of the accretion disc, implying a moderately high inclination.  

From {\it V}-band observations of 
the optical counterpart, \citet{motch} determined an optical ephemeris where 
minimum light occurs at 
{\it $T_{min}$} = JD 2,445,735.693 (4)  + 0.163890 (9) 
days, equivalent to a period of {\it $P_{opt}$} = 3.9334 $\pm$ 0.0002 hr.  
This broad optical modulation is probably due to the changing visibility of 
the heated face of the secondary star in the system, with an additional 
contribution from the X-ray heated bulge, although other explanations such as 
an asymmetrical disc or occultation of the disc by the companion cannot be 
ruled out.  
Within the errors, the optical and X-ray period measurements are consistent, 
with later X-ray observations indicating that the optical modulation provides 
a more accurate measurement of the orbital period \citep{smale2,levine}.  
A short 
section of simultaneous X-ray and optical coverage 
showed that the optical minimum occurs $\sim$ 0.16 in phase after the centre 
of the X-ray dips \citep{motch}.

Simple geometrical modelling is possible by considering the X-ray illuminated 
area of the companion star, disc radius, disc opening angle, and inclination 
at given distances which are compatible with the observed optical flux and the 
amplitude of the light curve.  This leads to constraints 
on the source inclination of 65$\degr$--73$\degr$, and on the distance of 
8-15 kpc \citep{motch}.

The first mass constraints for this system were based upon a unique 
relation between the orbital period and the mass of the companion, assuming 
that the companion is a low mass, zero-age main sequence star filling its 
Roche lobe \citep{warn}.   Using this approximation, 
\citet{courv} estimate the mass of the companion to be in the region of 
$\sim~0.45 M_{\sun}$.  

The first kinematic mass constraints were proposed by \citet{cow}, who 
obtained intermediate resolution ($\sim$ 3\AA) spectra using the 4m telescope 
at the Cerro Tololo Inter-American Observatory.  Assuming that velocities 
derived from the He\,{\sc ii}~$\lambda$4686 emission line centroid represent 
the motion of the compact star and not, for 
example, the streaming of gas between the stars or out of the system, then 
approximate stellar masses can be derived using the mass function.  The result 
is consistent with a low-mass main-sequence star ({\it M} $\simeq$ 0.5-0.8 
$M_{\sun}$) and a canonical 1.4 $M_{\sun}$ neutron star.

In the remainder of this paper we will attempt to confirm and better 
constrain these mass estimates using high-resolution blue optical 
spectroscopy.  We will measure the wings 
of the He\,{\sc ii}~$\lambda$4686 emission line to gain an estimate for the 
compact object velocity (Section 3.3), and use the Bowen Fluorescence 
technique \citep{SnC} to measure the donor star velocity (Section 3.4). 
We also obtained simultaneous X-ray observations (see Section 3.1) in order 
to re-establish 
the ephemeris of \citet{motch}.  This cannot be extended to the observations 
described herein because the propagated uncertainty in phase is now comparable 
to the orbital period itself.

\section{Observations}

We have obtained simultaneous optical and X-ray data for 
GR\,Mus / XB\,1254--690.  
The journal of observations is given in  Table \ref{tabobsmus}.

\begin{table}
\centering
\caption{Log of observations of GR\,Mus / XB\,1254--690}
\label{tabobsmus}
\scriptsize
\begin{tabular}{lccc}
\hline
\hline
HJD    & Observatory & Total Int. Time \\
\hline
2453151.52 & VLT & 3.75 h \\
2453151.51 & RXTE & 4.75 h \\
2453152.52 & VLT & 4 h \\
2453152.69 & RXTE & 5.25 h \\
2453153.53 & VLT & 2 h \\
\hline
\end{tabular}
\end{table}

\subsection{Optical data}

Phase-resolved spectroscopy of GR\,Mus was obtained from 26 to 28 May 
2004 using the FORS\,2 spectrograph mounted on the
VLT/UT4 at ESO's Paranal observatory.  During each night we 
observed GR\,Mus for $\simeq$ one full orbit, resulting in a total of 38 
spectra with an integration time of 900s each.  We used the 1400V 
volume-phased holographic grism, and a slit width of 0.7$''$, giving a 
wavelength coverage of
$\lambda$$\lambda$4513-5814 with a resolution of 70 km\,s$^{-1}$
(FWHM). The seeing during these observations varied between 0.5 and
2.1 arcsec. The slit was orientated at a position angle of 88.15$\degr$ to 
include a comparison star in order to calibrate slit losses.  We observed 
the flux standard Feige 110 with the same instrumental set-up in order to 
correct for the instrumental response.  He, Ne, Hg and Cd arc lamp exposures 
were taken during daytime 
for the wavelength calibration scale.  We de-biased and flat-fielded 
all of the images and used optimal extraction techniques to maximise the 
signal-to-noise ratio of the extracted spectra \citep{Horne86}.  
The pixel-to-wavelength scale was derived through polynomial fits to a large 
number of identified reference lines resulting in a dispersion of 0.64 
\AA\ pix$^{-1}$. 
Velocity drifts due to 
instrumental flexure (found to be always $<$5 km\,s$^{-1}$) were corrected by 
cross-correlation with the sky spectra.



\subsection{X-ray data}

Simultaneous X-ray data from XB\,1254--690 were obtained using the 
Proportional Counter Array (PCA) on board the {\it Rossi X-Ray Timing 
Explorer} ({\it RXTE}).  The Proportional Counter Array (PCA; for a detailed 
description, see \citet{jaho}) onboard the {\it RXTE} satellite consists of 
an array of 5 co-aligned
Proportional Counter Units (PCUs) that are sensitive to photons of
energy 2 to 60 keV with a total collecting area of 6500~cm$^2$.  We obtained 
36 ks of total coverage during our optical observations.  For our analysis we 
only used the data from the Standard 2 configuration, with a time resolution 
of 16 seconds.

\section{Analysis}

\subsection{Determination of orbital phase}

\begin{figure}
\includegraphics[height=1.0\linewidth,clip,angle=-90]{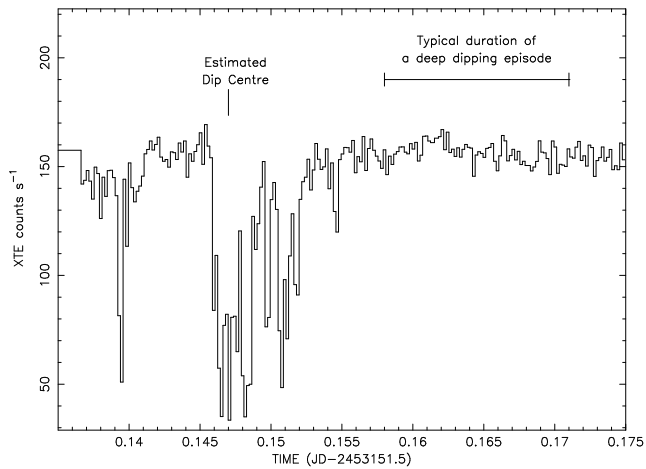}
 \caption{2--20 keV X-ray lightcurve showing the deep dip used to calibrate a 
new ephemeris for the system.  The dipping episode appears to begin shortly 
before JD\,2453151.64, and occurs for the typical dip duration of $\sim$1100 
seconds \protect\citep{smale2}.}
 \label{xray}
\end{figure}

The ephemeris of \citet{motch} is now out of date because the propagated 
uncertainty in phase is over a factor of 2 greater than the orbital period.  
We therefore 
need to make use of our contemporaneous {\it RXTE} observations to set a new 
zero point using the X-ray dips (note that we still use the \citet{motch} 
orbital period).  This is not quite as 
simple as it would first appear since there are large variations in dip depth 
and structure that are apparent from observation to observation, as well as 
cycle to cycle \citep{courv}.  Additionally, a note of caution must be sounded 
since in a number of earlier {\it RXTE} and {\it BeppoSAX} 
observations \citep{SnW,iaria}, X-ray dipping was not observed at all, with an 
accompanying decrease in the mean optical variability of the source.  With the 
combination of these effects, it was suggested that the vertical structure on 
the 
accretion disc edge, associated with the impact point of the accretion stream, 
had decreased in angular size at the time of these earlier observations.  The 
cause 
of this intermittent reduction in opening angle is not understood.  Our 
observations include just one deep dipping episode (Fig. \ref{xray}), 
co-incident with our first night of optical observations.

\begin{figure}
\includegraphics[height=1.0\linewidth,clip,angle=-90]{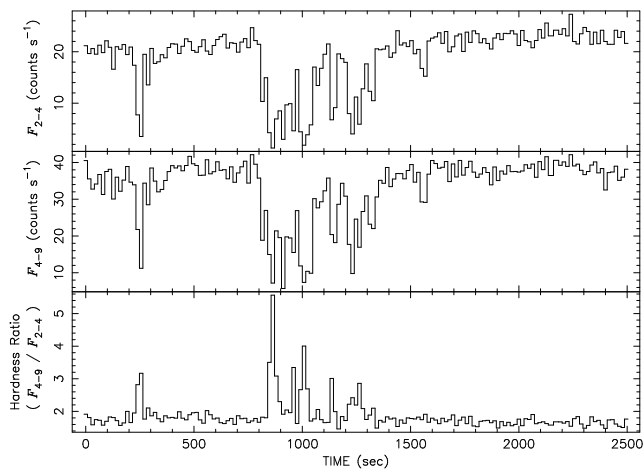}
 \caption{2--4 keV (top) and 4--9 keV (centre) X-ray lightcurves displaying 
the deep dipping episode.  The bottom panel displays hardness as a function 
of intensity.  There is an anti-correlation between intensity and hardness 
ratio during the deep dipping episode, which begins shortly after 200 seconds 
on the above scale.}
 \label{hard}
\end{figure}

XB\,1254--690 typically exhibits `shoulders' of dipping, i.e. 
regions both before and after the deep 
dipping in which there is a small decrease of intensity.  Each shoulder of the 
dip lasts $\sim$650--925 seconds, during which the total count rate in the 
2-25 keV band drops by $\sim$15\% with no strong changes in the hardness 
ratio.  The shoulder is followed by an interval of 
strongly variable deep dipping lasting $\sim$1100 seconds \citep{smale2}, with 
associated strong changes in the hardness ratio (Fig. \ref{hard}).  The 
hardening of the spectra in the dips is most likely to be due to absorption 
by cold matter.

Our 
observation of a dipping episode opens with the system already in the first 
shoulder phase, the onset of deep dipping occurring approximately 200--250 
seconds into the observation (before a brief return of the source intensity to 
the non-dip level), and then continuing to capture the full egress 
from deep dipping, into the shoulder and then returning to the continuum 
level.  Note that the return of the source intensity to the non-dip level 
during this episode is not unusual, and the light curve can in 
fact be interpreted as superpositions of small micro-dips \citep[e.g. ][]{uno}.

Unfortunately, the lack of information regarding the full 
ingress to deep dipping through the preceding shoulder introduces a degree of 
uncertainty into our estimate of the dip centre.  For example we could 
potentially be observing only a small section of an extended deep dipping 
episode \citep[e.g.][report dips lasting for $\sim$40 minutes]{uno}.  If this 
is the case, the absence of the beginning of the dip will cause the estimate 
for dip centre to be incorrect.  Nevertheless, we do have cause for optimism.  
The deep dipping which we observe is approximately 1100 seconds 
in length, equivalent to the typical period of a dipping episode in 
XB\,1254--690 (see above).  This is a strong indicator that we have full, or 
nearly-full, coverage of the deep dipping phase.  We estimate the dip centre 
by eye, taking 
the mid-point between the onset of and exit from deep dipping (which can be 
determined with the help of the associated hardness ratio changes, see Fig. 
\ref{hard}).  This occurs at $T_{dip}$ = JD 2453151.647 (3).  From 
\citet{motch} we know that the dip centre has an average orbital phase of 
0.84 (with phase 0 occurring at optical minimum).  We therefore estimate a new 
value for $T_0$ of JD 2453151.509 (3).

\begin{figure}
 \includegraphics[height=1.0\linewidth, clip,angle=-90]{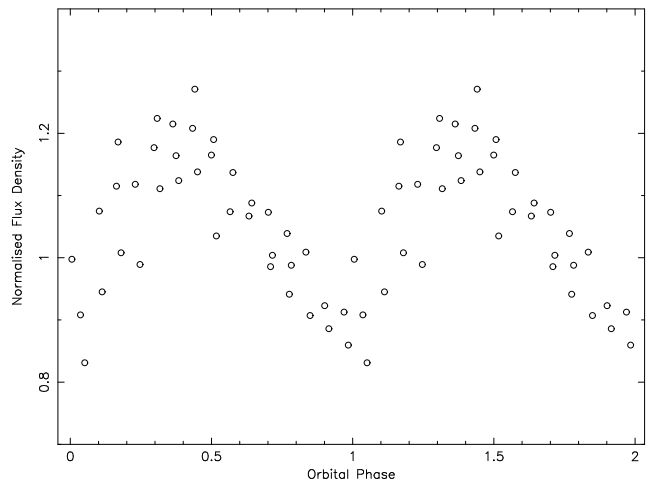}
 \caption{Continuum lightcurve of GR\,Mus, normalised to the flux level of 
the first observation in our dataset.  Note that the errors on each 
measurement are smaller than the plotted points.}
 \label{lcurve}
\end{figure}

There are problems with the reliability of an ephemeris generated from just a 
single dipping 
episode, not least because we do not have full coverage either side of the 
dip.  It is therefore hard to estimate the dip centre with absolute 
confidence.  However, we are encouraged by the fact that our dip lies 
within 0.05 phase of the 1999 dips  (the propagated phase uncertainty from 
these 
dips is of the order $\sim$0.6) reported by \cite{zand}.  More importantly, the 
continuum lightcurve from the resulting ephemeris possesses the same phasing 
relation reported by \cite{motch} (Fig. \ref{lcurve})\footnote{Note that the 
large intrinsic scatter in the optical continuum lightcurve makes the 
generation of a new ephemeris solely from a relatively limited dataset such as 
ours rather difficult.}.  In light of this 
evidence we are confident that we have identified the dip centre correctly 
and we will use this zero-point to calculate orbital phases throughout the 
remainder of this paper.

\subsection{Optical emission spectrum}

\begin{figure}
 \includegraphics[height=1.0\linewidth,clip,angle=-90]{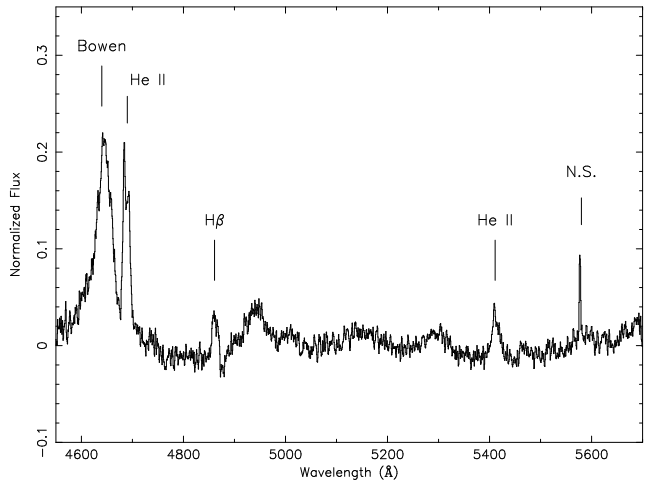}
 \caption{Average normalised spectrum of GR\,Mus.  The most prominent emission 
lines are indicated, including the Bowen blend of C\,{\sc iii} and N\,{\sc iii}, He\,{\sc ii} 
and H$\beta$.  The emission line near 5577\AA\ is a night sky feature.}
 \label{spec}
\end{figure}

The continuum subtracted average of the 38 individual optical spectra, 
corresponding to an exposure 
time of nearly 10 hours, is given in Fig. \ref{spec}.  The overall spectrum 
resembles that of other LMXBs, with high excitation lines superposed on a blue 
continuum.  This is typical of strongly X-ray heated accretion discs 
\citep[e.g. ][]{vPnMcC}.

The blue end of the optical spectrum in this system is dominated by the Bowen 
emission blend of C\,{\sc iii} and N\,{\sc iii}.  He\,{\sc ii} $\lambda$4686 
is also seen 
strongly in emission, in addition to weaker H$\beta$ and He\,{\sc ii} 
$\lambda$5411.  The absorption component longward of H$\beta$ is observed in a 
few other LMXBs such as X\,1822--371 \citep{cas}, XTE\,J2123--058 
\citep{hynes}, GX\,339--4 \citep{BnV} and MM\,Ser \citep{hynes2}.   

The narrow emission feature at 5577\AA\ is [O\,{\sc i}], a strong night-sky 
line useful for confirming the veracity 
of our wavelength calibration.  The broad features between 4900--5300\AA\ 
remain unidentified.  The absence of these features in our comparison star lead us to believe that they are real.  We also see some of these `bumps' in other LMXBs (e.g. GX\,9+9, 
X\,1822--371, V801\,Ara, V926\,Sco), and perhaps they are 
blends of a number of different weak emission lines.  Part of this could
be He\,{\sc i} $\lambda$4922 and $\lambda$5016, and also Fe\,{\sc ii}~
$\lambda$4924, 
$\lambda$5018, $\lambda$5169 and $\lambda$5297. There are also several 
O\,{\sc ii} and O\,{\sc iii} lines between 4891--5006\AA\ which may explain 
the large bump at $\sim$4930\AA.  Note also that the Bowen blend (and to a 
lesser extent the He\,{\sc ii} $\lambda$4686 emission line) appears to be 
sitting upon a broad emission `hump'.  This has been noted in other 
LMXBs \citep[e.g. V801\,Ara; ][]{august}, and attributed to a blend of 
Fe\,{\sc ii} emission lines \citep{schachter}.

\begin{figure*}
 \includegraphics[height=1.0\linewidth, clip,angle=-90]{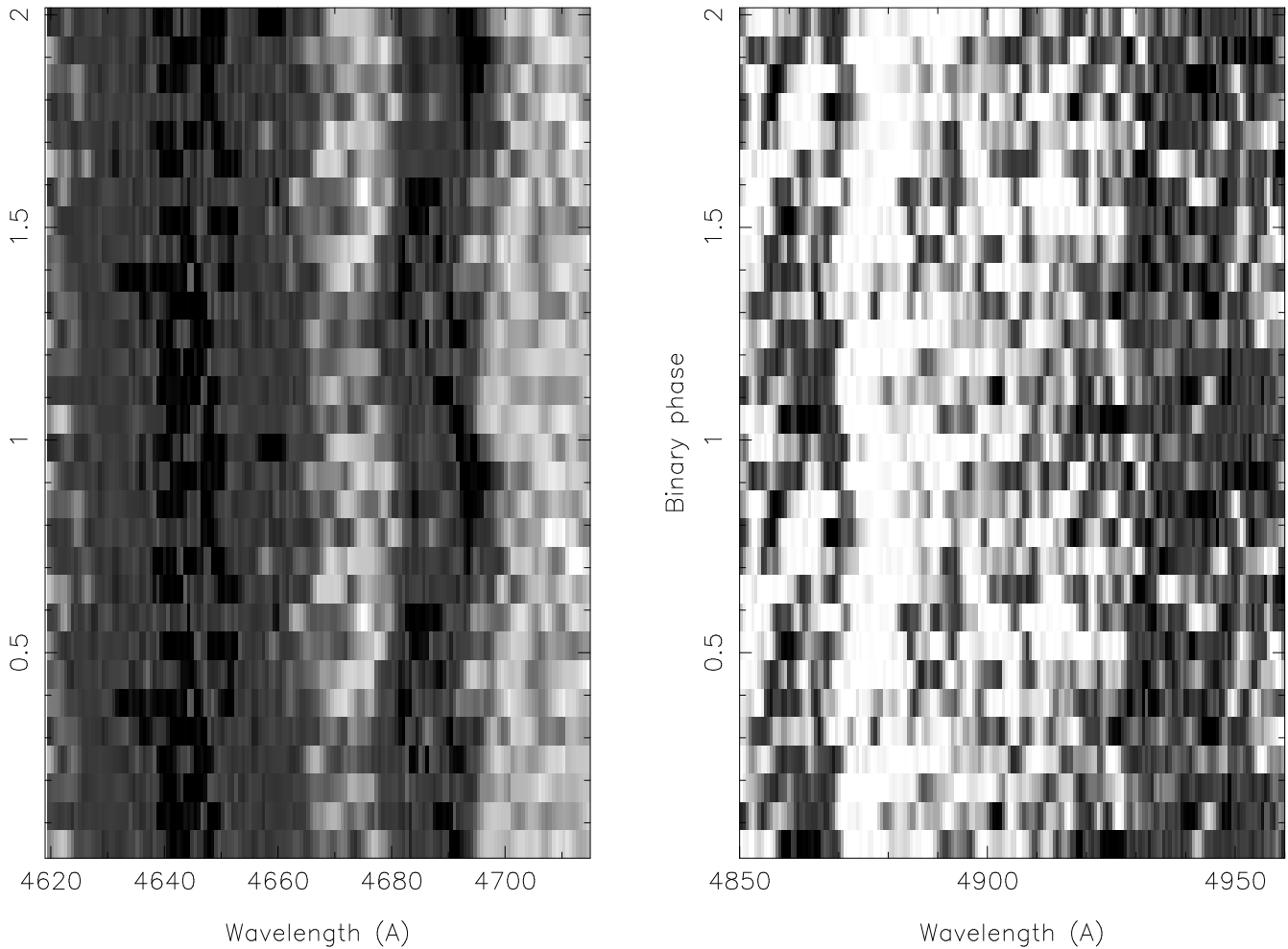}
 \caption{Trailed spectra of the Bowen complex and He\,{\sc ii} $\lambda$4686 
(left) and the H$\beta$ region (right).  The data are plotted twice for 
clarity.  Note that the wavelength scales differ.  There are clear multiple 
S-waves 
evident in the He\,{\sc ii} emission line, $\sim$0.25 out of phase with 
each other, and also in the Bowen complex.  The absorption component redwards 
of H$\beta$ is present at all phases.}
 \label{trail}
\end{figure*}

In Fig. \ref{trail}, we display the Bowen blend and He\,{\sc ii}~$\lambda$4686 
in 
the form of a trailed spectrogram.  All of the individual spectra were first 
normalized to the continuum using a third-order polynomial fit, and then 
phase-folded into 15 bins.  The spectra were then smoothed using a Gaussian 
filter with a FWHM of two pixels.  We see a 
clear S-wave structure modulated on the orbital period in the He\,II 
$\lambda$4686 emission.  The blue-to-red crossing phase is $\sim$0.6 and the 
semi-amplitude $\sim$250 km\,s$^{-1}$.  This behaviour indicates a close 
association with the accretion disc or compact object.  There is also an 
additional, weaker, component present.  Again, this 
is modulated on the orbital period, approximately a quarter cycle out of phase 
with the more prominent component.  There also appear to be faint traces of 
two anti-phased sharp components in the Bowen blend though the broad 
underlying component moves in phase with the accretion disc S-wave (maximum 
blue-shift at phase 0.25).  The tomograms discussed in Section 3.4 will 
help to characterise these spectral trails more thoroughly.

The H$\beta$ line is shown in the right-hand 
panel in Fig. \ref{trail}, along with the broad emission feature at 
$\sim$4930\AA .  The absorption component longward of H$\beta$ is present at 
all phases, and in common with the emission component displays an S-wave 
approximately in phase with the expected motion of the accretion disc.  The 
broad feature at $\sim$4930\AA\ also displays a S-wave.  The He\,II 
$\lambda$5411 emission line is 
too weak in the individual spectra to produce a trailed spectrogram.

\subsection{Motion of the compact object}

\begin{figure}
 \includegraphics[height=1.0\linewidth, clip,angle=-90]{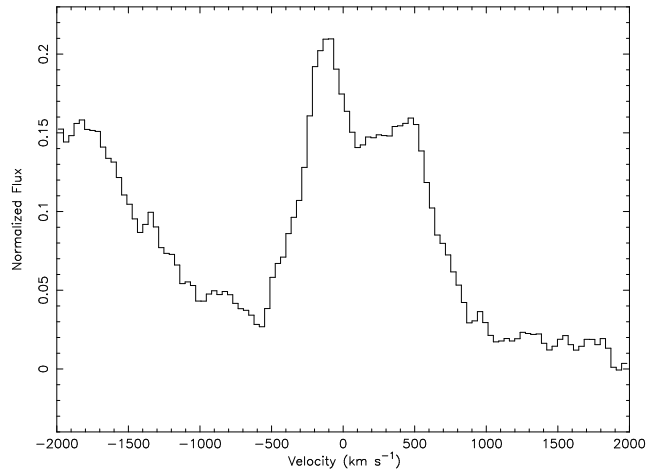}
 \caption{Velocity profile of the He\,{\sc ii} $\lambda$4686 emission line 
displaying a complex core structure.  The extreme blue wing is contaminated by 
the Bowen emission blend.}
 \label{profile}
\end{figure}

Early attempts to determine any kinematic constraints on this system met with 
limited success.  \citet{motch} noted that the relative weakness of the 
emission lines in their spectra did not allow for the study of their variation 
with orbital phase.  However, they did exclude the possibility of orbital 
motion of He\,{\sc ii} $\lambda$4686 (which requires a source of highly 
ionising radiation, and is typically seen in accretion discs) in excess of 300 
km\,s$^{-1}$.  Later observations by \citet{cow} did result in a measurement 
of the same He\,{\sc ii} line's core velocity, which possessed a 
semi-amplitude of 114 $\pm$ 13 
km\,s$^{-1}$.  However, this is not a particularly reliable measurement of the 
compact object velocity.  The line core can often be heavily contaminated by 
other emission sites (see Fig. \ref{profile}), for example from the accretion 
stream or `hot-spots' on the accretion disc itself.  

Instead, the compact 
object velocity can be better estimated by examining the wings of the emission 
line.  These wings find their 
origin in the high-velocity gas of the inner accretion disc, very close to the 
compact object itself.  Presumably this gas will closely replicate the 
motion of the compact object, and it is far less likely that alternative 
emission sites will provide a strong contaminating effect.  However, this 
method is not immune to contamination and we must be cautious in accepting the 
results of this analysis. 

\begin{figure}
 \includegraphics[height=1.0\linewidth, clip,angle=-90]{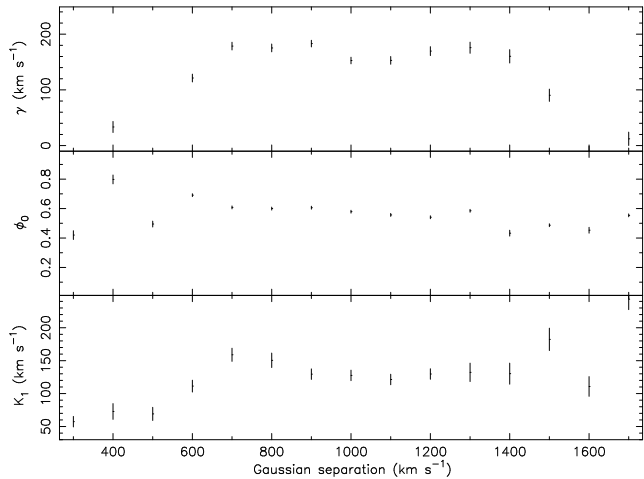}
 \caption{Diagnostic diagram for the He\,{\sc ii}~$\lambda$4686 emission line wings, 
displaying the systemic velocity $\gamma$ (top), phasing (centre) and radial 
velocity $K_1$ (bottom).  }
 \label{diag}
\end{figure}

We have followed the double-Gaussian technique \citep{SnY} to estimate both 
the velocity semi-amplitude $K_1$ and the systemic velocity $\gamma$ of the 
compact object from  He\,{\sc ii} $\lambda$4686.  We employed a double 
Gaussian 
bandpass with FWHM of 100 km\,s$^{-1}$ and Gaussian separations of 300--1700 
km\,s$^{-1}$ in steps of 
100 km\,s$^{-1}$. Radial velocity curves of line sections with separations of 
700--1400 km\,s$^{-1}$ yield consistent results with a median blue-to-red 
crossing phase of 0.58 $\pm$ 0.03, median velocity semi-amplitude $K_1$ = 130 
$\pm$ 16 km\,s$^{-1}$ and median systemic velocity $\gamma$ = 173 $\pm$ 12 
km\,s$^{-1}$ (Fig. \ref{diag}). The errors are conservatively taken from the 
maximum error obtained from sine wave fits to the results from each 
individual line section.  Separations below 700 km\,s$^{-1}$ suffer badly from 
contamination by the line core, whilst separations over 1400 km\,s$^{-1}$ 
become corrupted by continuum noise and the Bowen emission blend.  The small 
difference in phasing (0.08 orbits; slightly larger than the ephemeris 
uncertainty of $\sim$0.02 orbits) indicates that even the line wings may be 
contaminated to some extent by inhomogeneities in the disc.  This 
contamination will also affect the measurements of $K_1$ and $\gamma$ to some 
extent.  
Nevertheless, this remains 
the best method for estimating $K_1$ from 
our own and other currently available data.

Unfortunately, the He\,{\sc ii}~$\lambda$5411 and H$\beta$ emission lines, 
which are clearly seen in our averaged spectrum (Fig. \ref{spec}), are too 
weak in the individual spectra for a similar analysis.  The fitting procedures 
are simply dominated by noise.

\subsection{Doppler tomography}

We can use Doppler tomography \citep{MnH} to measure the radial velocity 
amplitudes of any weak, narrow, emission lines that may be present. By mapping 
the observed data onto a velocity coordinate frame, Doppler tomography makes 
use 
of all data at once and can thus be used to search for features that are too 
weak to be distinguished in each individual spectrum. One effectively resolves 
the 
distribution of line emission in the co-rotating frame of the binary system, 
providing an excellent tool for identifying the origin and kinematics of the 
various emission components. It also allows us to map the emission line 
distribution from the broader He\,{\sc ii} and H$\beta$ lines. 

\begin{figure}
 \includegraphics[height=1.0\linewidth, clip,angle=-90]{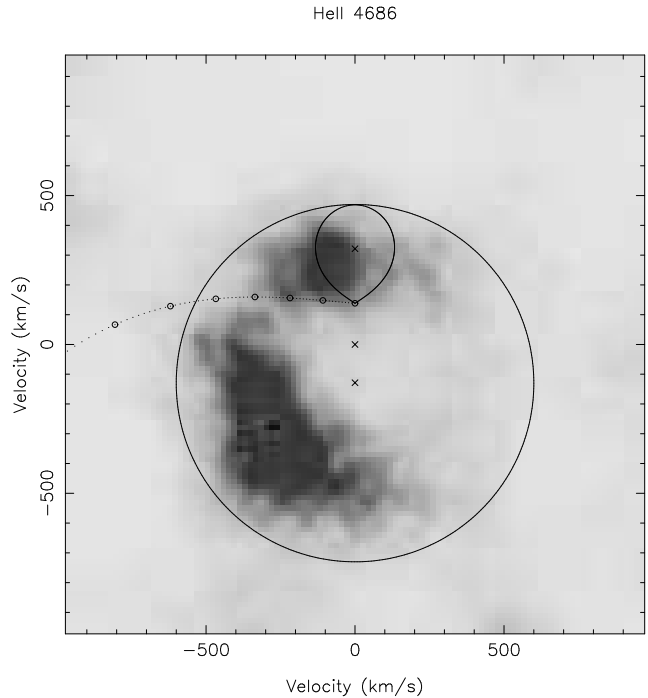}
 \caption{Doppler tomogram reconstructed from the He\,{\sc ii} $\lambda$4686 
emission line using maximum entropy optimisation, $\gamma$ = 185 km\,s$^{-1}$.  The area of the Roche lobe, 
the ballistic accretion stream and the 
centres of mass of the donor star (uppermost cross), system (map centre) and 
the compact object (lower cross) are plotted assuming a mass ratio of 
$\sim$0.4 and compact object velocity of 130 km\,s$^{-1}$.  The circle 
is centred upon the compact object and possesses a radius of 600 
km\,s$^{-1}$.}
 \label{dheii}
\end{figure}

\begin{figure}
 \includegraphics[height=1.0\linewidth, clip,angle=-90]{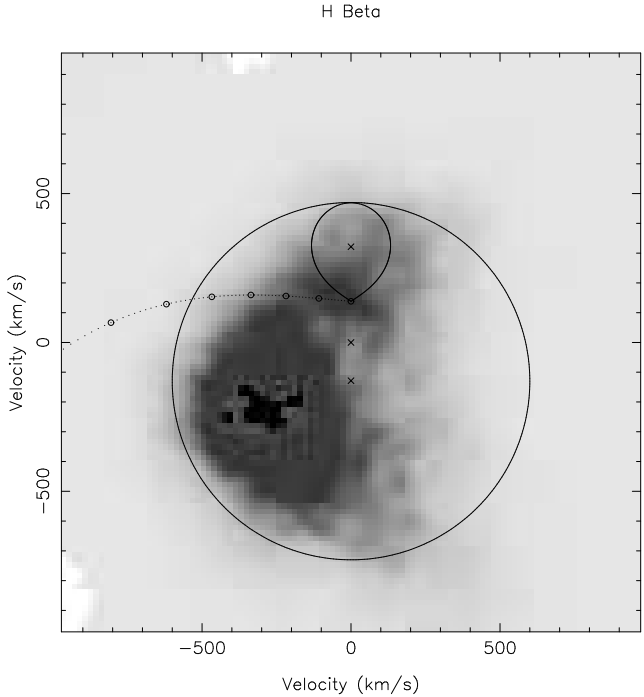}
 \caption{Doppler tomogram reconstructed from the H$\beta$ emission line 
using maximum entropy optimisation, $\gamma$ = 185 km\,s$^{-1}$.  The symbols are as for Fig. \ref{dheii}.}
 \label{dhb}
\end{figure}

\begin{figure}
 \includegraphics[height=1.0\linewidth, clip,angle=-90]{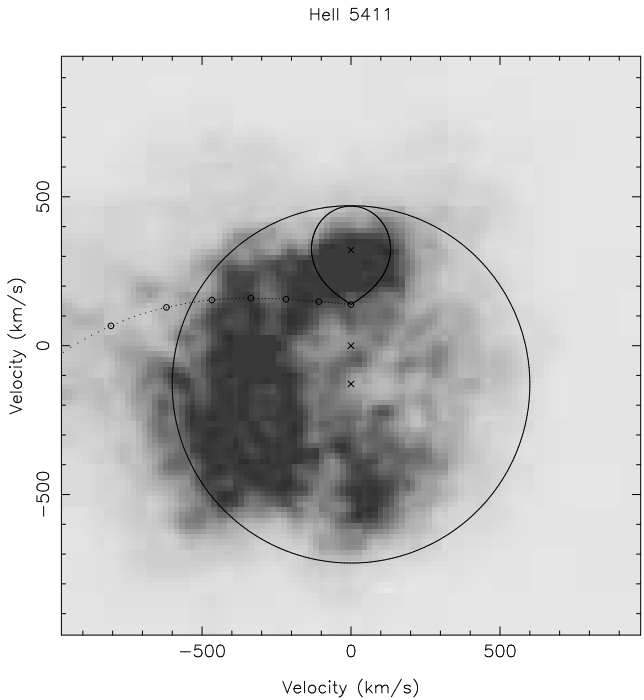}
 \caption{Doppler tomogram reconstructed from the He\,{\sc ii} $\lambda$5411 
emission line using maximum entropy optimisation, $\gamma$ = 185 km\,s$^{-1}$.  The symbols are as for Fig. 
\ref{dheii}.}
 \label{d5411}
\end{figure}

\begin{figure}
 \includegraphics[height=1.0\linewidth, clip,angle=-90]{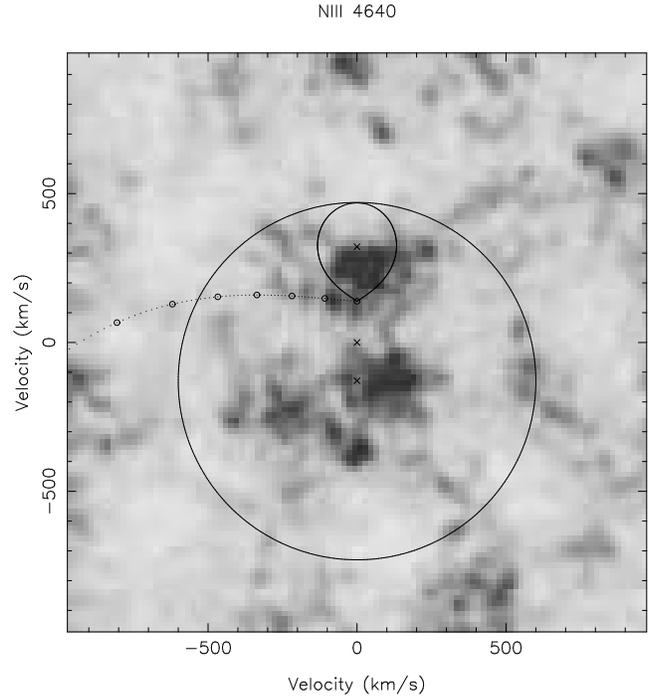}
 \caption{Doppler tomogram reconstructed from the N\,{\sc iii}~$\lambda$4640 
emission line using maximum entropy optimisation, $\gamma$ = 185 km\,s$^{-1}$.  The symbols are as for Fig. 
\ref{dheii}.}
 \label{dniii}
\end{figure}

Secondary star 
emission is readily identifiable in Doppler tomograms since the solid body 
rotation of the Roche lobe is mapped to a corresponding Roche lobe area along 
the positive V$_y$-axis. For reference, in Figs. \ref{dheii}--\ref{dniii} we 
overplot the area of the Roche lobe and the centres of mass of the donor 
star (uppermost cross), full system (map centre) and the compact object 
(lower cross) assuming a mass ratio of $\sim$0.4 \citep[equivalent to the mass 
ratio proposed by ][]{cow} and compact object velocity 
of 130 km\,s$^{-1}$.  Additionally, we have overplotted the ballistic 
accretion stream for such a system.  The circle, centred upon the 
compact object and with radius 600 km\,s$^{-1}$, indicates the common 
emission regions from a generic LMXB accretion disc.  We employed a maximum 
entropy implementation of Doppler tomography whereby the data are fitted under 
the added constraint of maximizing the entropy (i.e. image smoothness) of the 
tomogram \citep{MnH}. This reduces the presence of noise artifacts in the 
recovered tomograms and allows for a simultaneous fit to a number of heavily 
blended lines.

In Fig. \ref{dheii} we display a tomogram of the He\,{\sc ii}~$\lambda$4686 
emission 
line.  This displays a clear enhancement in the lower left quadrant and an 
additional component close to the accretion stream.  We do not observe the 
classic ring-like distribution of the accretion disc, presumably due to the 
strength of the line core features vastly overwhelming the high velocity inner 
disc emission.  For this reason we cannot be fully confident in the results of 
the double Gaussian technique (see Section 3.3), though this remains at present the best method for estimating $K_1$.  

The enhancement in the 
(lower) left quadrant is also seen in the tomograms of H$\beta$ 
(Fig. \ref{dhb}) and He\,{\sc ii} $\lambda$5411 (Fig. \ref{d5411}).  
Whilst this 
phenomenon is a common feature in the remarkable SW\,Sex-type systems 
\citep{hell}, it has 
also been noted in some other LMXBs, notably XTE\,J2123--058 \citep{hynes}, 
GX\,9+9 \citep{corn2} and J\,0422+32 \citep{cas1}.  However, in this 
case \citep[and in the case of X\,1636--536 and X\,1735--444, ][]{cassy} 
the emission is not quite so localised in the lower-left 
quadrant, but is present in a broad band on the left-hand side of the maps.  
Rather than being an analogue to the highly magnetic SW\,Sex-type systems, it 
is likely that this emission is produced in an extended disc bulge.  
In most LMXBs, the He\,{\sc ii} emission is predominantly produced in a 
hotspot where the accretion 
stream impacts the disc \citep[thus producing a spot in the upper left quadrant of a 
Doppler map, e.g. ][]{pear} or is more evenly distributed around the bulk of the accretion 
disc.  If the emitting gas was symmetrically distributed around the primary, 
the Doppler image would reveal a circular distribution centred on $V_x$ = 0, 
$V_y$ = -$K_1$.  This would have been a useful way of confirming the $K_1$ 
estimate derived from the wings of the He\,{\sc ii} 
$\lambda$4686 emission line.  

Like He\,{\sc ii}~$\lambda$4686, the He\,{\sc ii}~$\lambda$5411 
tomogram also indicates the presence of some diffuse emission at velocities 
which partially correspond to the Roche lobe of the donor star.  However, as 
this emission is diffuse in the Doppler map, extending over a broad range of 
velocities, it is difficult to directly
associate this extended emission with the donor star.  Fluorescent 
emission from the donor in this case should manifest itself as a tightly 
concentrated spot situated on the positive $V_y$ axis.

In Fig. \ref{dniii} we present a tomogram of the N\,{\sc iii} 
$\lambda$4640 emission 
line.  The maximum entropy method of producing Doppler tomograms allows for a 
simultaneous fit to a number of different lines to produce one map.  However, 
in this case the N\,{\sc iii} $\lambda$4640 line strongly dominates over any 
others in the Bowen blend.  We see two regions of enhanced emission.  The 
brighter of the two spots occurs at a combination of phasing and velocity 
where we would expect the donor star to be fluorescing.  The small horizontal 
offset of $V_x$$\simeq$35$\pm$30 km\,s$^{-1}$ is consistent with the errors in 
our ephemeris ($\pm$ 0.02 in phase).  

The second spot occurs at a much lower 
velocity, and is 1--2 
$\sigma$ weaker (relative to the local noise level) than the first.  It is 
difficult to provide a truly 
satisfactory explanation for its origin.  Perhaps it traces some 
form of low velocity outflow (e.g. disc wind), though why it should be 
preferentially observed at an apparent phase of $\sim$0.25 remains a mystery 
(orbital phase can be considered to increase 
in a clockwise-fashion around a Doppler map).

\begin{figure}
 \includegraphics[height=1.0\linewidth, clip,angle=-90]{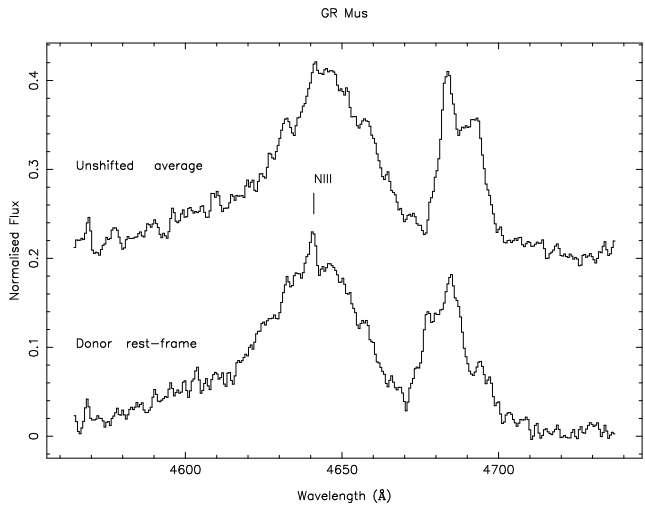}
 \caption{Average spectrum created by combining all of the individual 
unshifted spectra from our observations (top), showing the Bowen emission 
blend and He\,{\sc ii} $\lambda$4686 region.  Below this is shown the average 
spectrum when all of the individual spectra are shifted into the rest-frame of 
the donor star.  A sharp N\,{\sc iii} $\lambda$4640 component is clearly 
resolved.}
 \label{donorspec}
\end{figure}

If we presume that the bright spot in Fig. \ref{dniii} is caused by the 
irradiated inner face of the donor star, this is an excellent tracer for the 
kinematics of the system.  The precise location of the spot can be measured 
using a two-dimensional Gaussian fit (though bearing in mind that the actual 
shape of the spot is not expected to be Gaussian), providing a velocity of 
$K_{\rm em}$ = 245 $\pm$ 30 km\,s$^{-1}$.  Using this value we can shift all 
of our spectra into the rest-frame of the (irradiated face of the) donor.  A 
sharp N\,{\sc iii}~$\lambda$4640 component is clearly resolved (Fig. 
\ref{donorspec}).  However, since this is a measure of the motion of the 
irradiated inner face of the 
donor star, it only provides us with a lower-limit on the velocity of the 
centre of mass \citep[see e.g.][]{SnC,teo}.


One final piece of information remains to be extracted from the Doppler 
tomograms;  assigning the correct systemic velocity should produce the most 
sharply defined features in a Doppler map.  We used the He\,{\sc ii} and 
N\,{\sc iii} 
Doppler maps to verify the systemic velocity.  The $\chi^2$ value of the map 
was calculated for a range of $\gamma$, and the best fit in terms of minimal 
$\chi^2$ was achieved for $\gamma$ = 192 $\pm$ 2 km s$^{-1}$ for the 
N\,{\sc iii} 
map and $\gamma$ = 185 $\pm$ 2 km s$^{-1}$ for He\,{\sc ii}~$\lambda$4686.  
These are both consistent with, if slightly higher than, the results derived 
from the He\,{\sc ii} $\lambda$4686 wings in Section 3.3.  The complex, 
asymmetric core profile of the He\,{\sc ii} emission line could be 
contaminating one wing more 
than the other, leading to this small offset.  We heretofore adopt the 
value for $\gamma$ derived from the He\,{\sc ii}~$\lambda$4686 Doppler map.  
This should be more reliable than the N\,{\sc iii} map since only a single, 
isolated emission line is involved in the fitting process, rather than the 
more complicated blend of Bowen lines.

\section{Discussion}

\subsection{Systemic velocity}

Whilst the systemic velocities inferred from both the wings of the 
He\,{\sc ii} 
$\lambda$4686 line and the Doppler tomograms are self-consistent ($\sim$ 
185 km\,s$^{-1}$), they are considerably higher than the earlier estimate by 
\citet{cow} (98 $\pm$ 10 km\,s$^{-1}$).  This measurement was obtained from 
rather low quality data, with poor spectral and temporal resolution.  
Additionally, they based their measurements on the motion of the He\,{\sc ii} 
$\lambda$4686 line core.  There are a number of different emitting regions 
which contribute to this line, leading to a complex core structure (see e.g. 
Figs. \ref{profile}, \ref{dheii}).  A basic measurement of the line centre 
under the assumption that this will accurately trace the motion of the compact 
object will therefore be subject to large systematic errors.  

In light of 
this, we believe that our estimate for the systemic velocity will be a 
better reflection of the true value.  Although large, our measurements do seem 
reasonable in the context of the high galactic latitude ($b$ = -6.42) of 
GR\,Mus; the neutron star is likely to have received a substantial 
`kick' out of the Galactic Plane at birth.  A number of statistical studies 
on pulsar velocities have been carried out.  These studies give a mean birth 
velocity of 100--500 km s$^{-1}$, with some
possessing velocities of $\ge$1000 km s$^{-1}$ \citep{wang}.

\subsection{Interpretation of the Bowen blend Doppler map}

The N\,{\sc iii} Doppler map (Fig. \ref{dniii}) appears to display enhanced 
emission in the region of and at the phasing where we would expect to see the 
companion star Roche lobe, in addition to an 
abundance of noisy structure in the outer regions of the map.  A useful 
way to confirm the validity of the Doppler map is to observe the behaviour 
of sharp components in our trailed spectrogram (Fig. \ref{trail}).  This is 
not trivial due to the faintness of GR\,Mus compared to a 
system such as Sco\,X-1, which clearly features narrow Bowen lines moving in 
phase with the donor star \citep{SnC}.

\begin{figure}
 \includegraphics[width=1.0\linewidth, clip,angle=180]{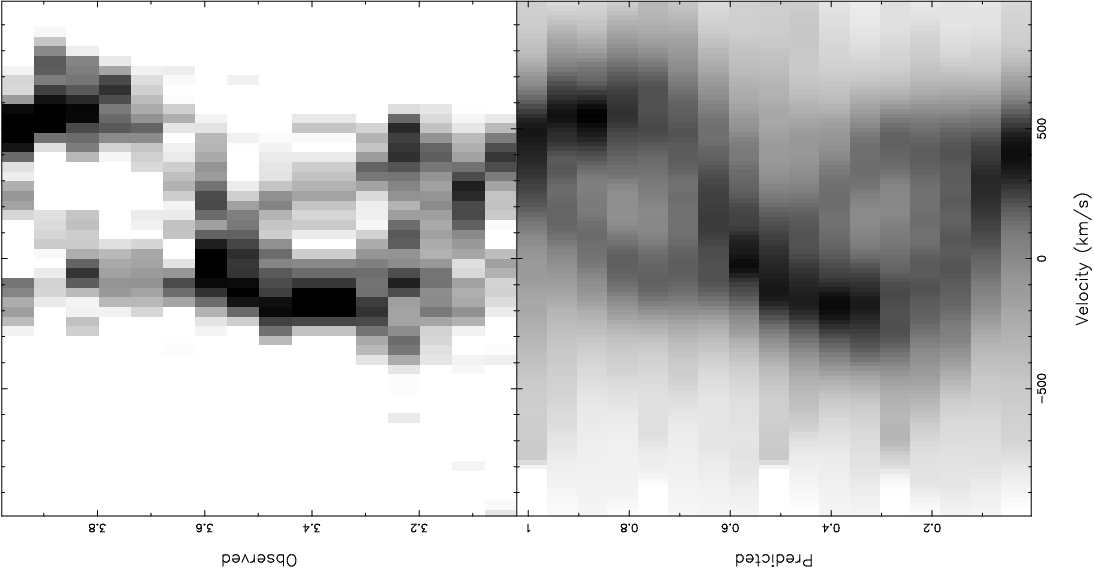}
 \caption{The right hand panel displays the observed trailed spectra of the 
He\,{\sc ii} $\lambda$4686 emission line and the left hand panel shows the 
ideal 
trailed spectrogram reconstructed from the Doppler tomogram shown in Fig. 
\ref{dheii}.  Orbital phase is plotted along the $x$-axis.}
 \label{pheii}
\end{figure}

\begin{figure}
 \includegraphics[width=1.0\linewidth, clip,angle=180]{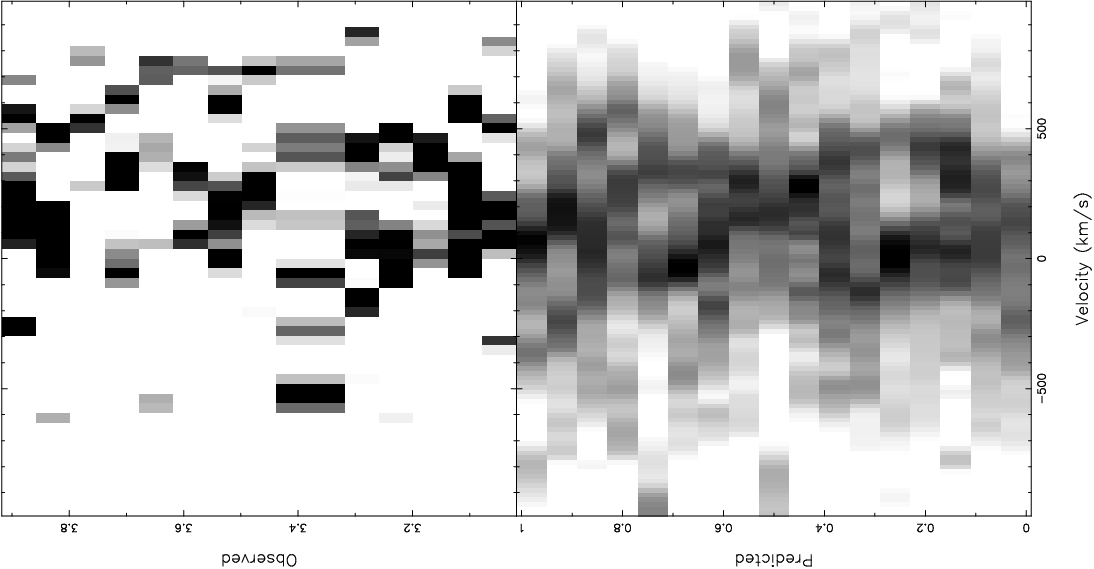}
 \caption{The right hand panel displays the observed trailed spectra of the 
N\,{\sc iii} $\lambda$4640 emission line, part of the Bowen emission complex.  
Next to this is plotted the ideal trailed spectrogram reconstructed from the 
Doppler tomogram in Fig. \ref{dniii}. Orbital phase is plotted along the 
$x$-axis. }
 \label{pniii}
\end{figure}

We can reconstruct an ideal trailed spectrogram from the Doppler tomograms 
to test for systematic residuals in the maps.  
As a test, we first apply this technique to the He\,{\sc ii} 
$\lambda$4686 emission 
line.  The two distinct components we see in the Doppler map (Fig. 
\ref{dheii}) are reproduced clearly in the predicted trailed spectrogram 
(left hand panel, (Fig. \ref{pheii}), and can be easily identified in the 
observed data.  For the case of the N\,{\sc iii} $\lambda$4640 emission line 
(Fig. \ref{pniii}) we again see two components in the reconstructed 
spectrogram, though not quite so distinctly as for the case of He\,{\sc ii}.  
Comparing this 
idealised image to the real data it is indeed possible to detect faint traces 
of these two narrow, approximately anti-phased lines which produce the two 
distinct spots in Fig. \ref{dniii}.  The second spot is therefore unlikely to 
be a systematic residual produced when creating the Doppler map, and its 
unusual phasing/positioning could perhaps therefore be attributed to a region 
which violates one of the axioms of Doppler Tomography, for 
example that all motion should be parallel to the orbital plane.  Note that 
a violation of these axioms does not mean that the map is erroneous, but it 
does complicate the successful interpretation of the map.

\subsection{Refined mass limits}


\cite{jong} estimated an opening (semi-) angle for the accretion disc in 
GR\,Mus of $\sim 12\degr$, using a simple geometric model for the 
reprocessing of X-rays in LMXBs.  It is important to note that this value is 
hostage to the implicit assumptions underlying these models, for example the 
geometric description of the disc and of the X-ray emitter.  This will entail 
a systematic uncertainty that likely exceeds the internal model accuracy of 
less than a degree.  Nevertheless, this value agrees well with \cite{motch} 
who estimated an opening angle of $\sim 9\degr-13\degr$ in order to explain 
the observed amplitude of the orbital photometric modulation.  

We can use this information 
to derive a mass estimate for the individual components by 
using the `K-correction' algorithm developed by \cite{teo}.  This measures the 
deviation 
between the reprocessed light centre of the fluorescence emission lines and 
the centre of mass of a Roche lobe filling star in a persistent LMXB, 
including the screening effects by a flared accretion disk.  Using the fourth 
order polynomial fits given by \cite{teo} for disc opening angles of 8$\degr$, 
10$\degr$, 12$\degr$ and 14$\degr$ gives a total mass range for the neutron 
star of 1.20 $\leq M_X/M_{\sun} \leq$ 2.64.  Since the neutron star mass and 
the mass ratio are related by the mass function\footnote{$f(M_X) = \frac{K_2^3 P_{orb}}{2\pi G} = 
\frac{M_X \sin ^3i}{(1 + q)^2}$}, we obtain 0.32 $\leq q \leq$ 
0.43, and limits on the companion mass of 0.45 $\leq M_2/M_{\sun} \leq$ 0.85.  

This correction is dependent upon the opening angle of the 
accretion disc, monotonically decreasing with opening angle such that the 
largest opening angle gives the lowest neutron star mass.  Since the largest 
opening angle we have used exceeds the estimates of both \cite{motch} and 
\cite{teo}, this gives a very conservative lower limit to the neutron star 
mass, and the true value is unlikely to be so low.  Equally, the upper limit 
is derived using an opening angle lower than that suggested by \citet{motch} 
and as such the neutron star is unlikely to be quite so massive.

\begin{table}
\centering
\caption{Derived system parameters for GR\,Mus / XB\,1254--690}
\label{tabparmus}
\scriptsize
\begin{tabular}{lclc}
\hline
\hline
Parameter  & & Parameter & \\
\hline
$T_{dip}$ (HJD) & 2453151.647 (3) & $\gamma$ (km\,s$^{-1}$) & 185 $\pm$ 2 \\
$T_0$ (HJD) & 2453151.509 (3) & $q$ & 0.32--0.43 \\
$K_1$ (km\,s$^{-1}$) & 130 $\pm$ 16 & $M_X (M_{\sun})$ & 1.20--2.64 \\
$K_{\rm em}$ (km\,s$^{-1}$) & 245 $\pm$ 30 & $M_2 (M_{\sun})$ & 0.45--0.85 \\
\hline
\multicolumn{4}{l}{Note: Inclination constrained to 65$\degr$--73$\degr$; \citet{motch}}
\end{tabular}
\end{table}

Limiting ourselves to a disc opening angle of exactly 12$\degr$ \citep{jong}, 
the mass constraints upon the neutron star tighten up to 
1.35 $\leq M_X/M_{\sun} \leq$ 2.32, with 0.33 $\leq q \leq$ 0.41 and 
0.49 $\leq M_2/M_{\sun} \leq$ 0.79.  These limits are not a vast improvement 
over the ones derived above using a range of opening angles from 
8--14$\degr$.  This indicates that the major source of uncertainty in this 
case is still incurred from the measurements of $K_{\rm em}$ and particularly 
of $K_1$ (see Section 3.3), in 
addition to the system inclination angle.  The choice of disc opening angle 
(within a reasonable range) has only a small effect in comparison.  
Nevertheless, we feel it is prudent not to restrict our final mass estimate 
by the assumption of a disc opening angle of exactly 12$\degr$, 
preferring the more conservative range of 8--14$\degr$.  The derived 
parameters for GR\,Mus are listed in Table \ref{tabparmus}.

\subsection{Roche lobe size}

We know that the 
companion must be filling its Roche lobe in order to permit persistent 
accretion.  There is a well known and useful 
relationship between the fractional Roche lobe radius and the mass-ratio 
\citep{pac} which can be combined with Kepler's 3rd 
law to 
provide an estimate of the size of the Roche lobe as a function of period.  
In the case of this system (with a 3.9 hour period), the relationship may be 
expressed as:  $R_L = 0.58 M_2^{1/3}$.

We can thus derive the range of potential Roche lobe radii 
consistent with the mass range of the companion star that we have derived 
kinematically.  This turns out to be 0.44 $\leq R_L/R_{\sun} \leq$ 0.55.  Whilst at the 
lower end of our mass constraints the donor would be consistent with a 
main-sequence-like object as proposed by \citet{motch}, we cannot rule out the 
possibility that GR\,Mus harbours an over-massive evolved companion 
that has been stripped of its outer layers \citep[e.g.][]{kingschenk}.

\section{Conclusions}

We have used the Bowen emission blend and Doppler tomography to trace for the 
first time the motion of the donor star in the persistent LMXB GR\,Mus.  
This has allowed us to derive kinematical mass limits for this 
system.  In combination with estimates for the opening angle of the 
accretion disc, and new measurements of the He\,{\sc ii}~$\lambda$4686 
emission line 
(in an attempt to trace the motion of the compact object), we have derived 
tentative new mass 
constraints of 1.20 $\leq M_X/M_{\sun} \leq$ 2.64 for the compact object and 
0.45 $\leq M_2/M_{\sun} \leq$ 0.85 for the companion star.  

We cannot rule out the possibility that GR\,Mus harbours an over-massive 
evolved companion star, or more intriguingly an over-massive neutron star.  
Tighter constraints on the system parameters may therefore have important 
implications for our knowledge of the equation of state of nuclear matter 
\citep[e.g.][]{COOK} or indeed on the formation scenarios for typical LMXBs 
\citep[e.g. ][]{pfahl}.

Future investigations should also concentrate on determining the origin of 
the unusual second spot in the N\,{\sc iii} Doppler map (Fig. \ref{dniii}).  
This feature is hard to explain, and perhaps could be caused by a violation 
of the axioms of Doppler Tomography.  Related to this point, the 
broad emission band seen in the left hand side of the Doppler maps for 
He\,{\sc ii} and H$\beta$ (in this system and a number of others) is of 
uncertain origin.  We believe this emission to be produced in an extended 
disc bulge, but physical modelling will be necessary to confirm this.

\section*{Acknowledgments}

ADB was supported by a PPARC Research Studentship.  DS acknowledges 
a Smithsonian Astrophysical Observatory Clay Fellowship as well as support 
through NASA GO grant NNG06GC05G.  Based on 
observations made with ESO Telescopes at the Paranal Observatory under 
programme ID 073.D-0819.  
We gratefully acknowledge the use of the {\sc molly} and {\sc doppler} 
packages developed by T.R. Marsh.

\label{lastpage}

\end{document}